\documentclass[lettersize,journal]{IEEEtran}
\IEEEoverridecommandlockouts
\usepackage{cite}
\usepackage{amssymb,amsfonts}
\usepackage{algorithm} 
\usepackage{graphicx}
\usepackage{optidef}
\usepackage{textcomp}
\usepackage{xcolor}
\usepackage{algpseudocode}

\usepackage{atbegshi}
\AtBeginDocument{\AtBeginShipoutNext{\AtBeginShipoutDiscard}}

\def\BibTeX{{\rm B\kern-.05em{\sc i\kern-.025em b}\kern-.08em
    T\kern-.1667em\lower.7ex\hbox{E}\kern-.125emX}}

\begin{document}

\title{A Novel Deep Reinforcement Learning-based Approach for Enhancing Spectral Efficiency of IRS-assisted Wireless Systems
}

\author{Farimehr Zohari, S. M. Mahdi Shahabi and Mehrdad Ardebilipour}

\thanks{Farimehr Zohari and Mehrdad Ardebilipour are with the Department of Electrical Engineering, K. N. Toosi University of Technology, Iran. (e-mail: Farimehr.Zohari@email.kntu.ac.ir;  mehrdad@eetd.kntu.ac.ir).} 
\thanks{S. M. Mahdi Shahabi is with the Department of Engineering, Kings College London, U.K. (e-mail: mahdi.shahabi@kcl.ac.uk).}
\thispagestyle{empty}
\clearpage
\pagenumbering{arabic} 
\maketitle

\begin{abstract}
This letter investigates an intelligent reflecting surfaces  (IRS)-enhanced network from spectral efficiency enhancement point of view for downlink multi-user (MU) multi-input-single-output systems (MISO).  In contrast to previous works which mainly focused on alternative optimization methods, we investigate the non-convex joint optimization problem of the active transmit beamforming matrix at the base station together with the passive phase shift matrix at the IRS by utilizing two deep reinforcement learning frameworks, i. e., deep deterministic policy gradient (DDPG) and twin delayed DDPG (TD3). Simulation results reveal that the neural networks in the latter scheme perform generally more satisfactorily in various situations.
\end{abstract}


\section{Introduction}
\IEEEPARstart{I}{ntelligent} reflecting surfaces (IRSs) have been considered as a promising technology for achieving the expected spectral efficiency (SE) and energy efficiency (EE) as well as the cost efficiency for beyond-fifth-generation (B5G) wireless communication in the recent research \cite{ref1}. By compensating for the power loss over long distances, IRSs are able to modify the wireless propagation environment. Thanks to passively reflecting the radio signals that are impinging, base stations (BSs) and users are able to create virtual line-of-sight (LoS) relationships, which might potentially improve the received signal-to-interference-plus-noise ratio (SINR) \cite{ref2}. The IRS is a two-dimensional (2D) electromagnetic (EM) material surface, referred to as a metasurface made up of a wide variety of passive scattering elements with a unique physical structure. In order to alter the EM properties, e. g. the phase shifts of the reflection of the incident RF signals upon the scattering elements, each scattering element might be controlled in a software-defined manner. The reflecting phases and angles of the incident RF signals can be freely modified to provide a desired multi-path effect via a joint phase control of all scattering components \cite{ref2}. 

In order to enhance the communication performance, transmit beamforming at the BS and passive beamforming at the IRS should be cooperatively constructed \cite{ref2}. Extensive studies have been done by various researchers to solve the non-convex joint optimization problem. In \cite{ref4}, the focus was on joint transmit beamforming and phase shift of the IRS in multiple-input-multiple-output (MIMO) systems in order  to enhance users fairness based on a number of alternative optimization techniques. Authors in \cite{ref5} investigated the non-trivial tradeoff between the EE and the SE in multiuser MIMO uplink communications with the use of an IRS outfitted with discrete phase shifters utilizing an iterative mean-square error minimization approach. In \cite{ref6}, a new deep reinforcement learning (DRL) framework was designed for the joint design of transmit beamforming matrix at the base station and the phase shift matrix at the IRS in a multiple-input-single-output (MISO) systems using a deterministic policy gradient (DDPG) method to increase the sum rate. In \cite{ref7}, the authors concentrated on machine learning (ML) approaches for performance maximization in IRS-assisted wireless networks.

In this letter, a new DRL algorithm called Twin Delayed DDPG (TD3) is employed so as to jointly design transmit beamforming at the BS and phase shifts at the RIS in order to improve SE in downlink multi-user (MU) MISO systems, whereas the vast majority of previous works utilized alternative optimization algorithm dealing with high mathematical complexity levels. The direct channels between the BS and the users are assumed to be hardly ever blocked by any obstacles, and consequently are considered in the problem formulation, in addition to  assuming the global \emph{channel state information} (CSI) available at both the IRS and BS. Specifically speaking, this method has not been utilized in any work prior to this study in this system model. In this regard, first, the desired system model is characterized and the mathematical optimization equation of the SE for our system model is derived. Then, the structures of TD3 framework will be elaborated and its differences with DDPG will be highlighted. The result of simulations reveals this novel DRL framework shows notably better performance compared to DDPG.
\section{ System Model and Problem Formulation}

The considered MISO system consists of a BS which is equipped with $M$ antennas, an IRS that has $N$ reflecting elements, and $K$ single-antenna users (Fig. 1). The channel matrix between the BS and the IRS is assumed $\textbf{H}_1 \in \mathbb{C}^{(N \times M)}$, the channel vector between the IRS and the \emph{k}-th user and the channel vector between the BS and the \emph{k}-th user are presumed $\textbf{h}_{r,k} \in \mathbb{C}^{(N \times 1)}$ and $\textbf{h}_{d,k} \in \mathbb{C}^{(M \times 1)}$, respectively, for $k \in[ 1, K]$.
\begin{figure}[htbp]
\centerline{\includegraphics[width=2.5 in]{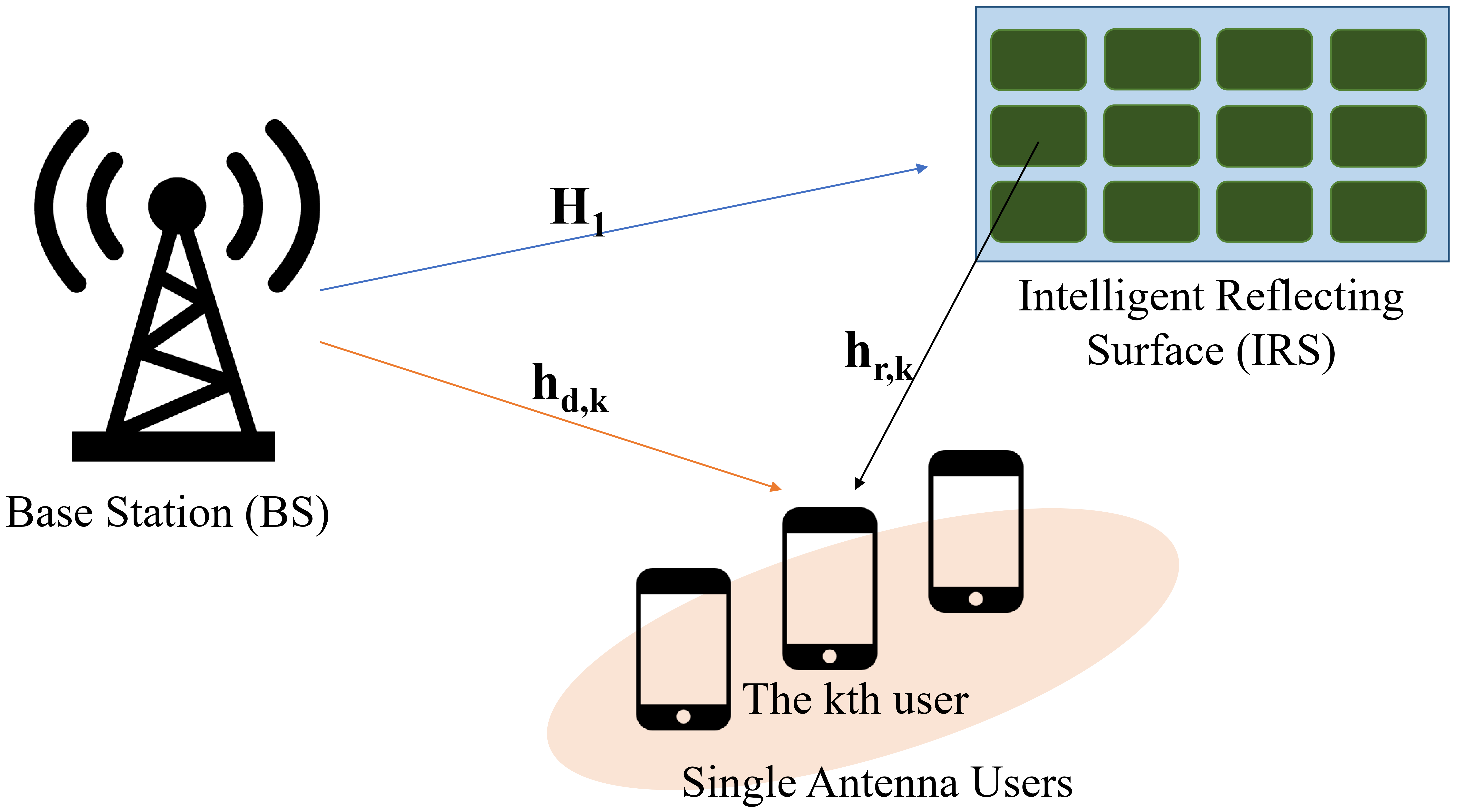}}
\caption{An IRS-aided multiuser MISO communication system.}
\label{fig}
\end{figure}
The received signal a the $k$-th user  can be represented as
\begin{equation}
    y_k = (\textbf{h}^T_{r,k}\Phi \textbf{H}_1 + \textbf{h}^T_{d,k})\textbf{G} \textbf{x} + \omega_k,
\end{equation}
where ${\Phi} = \text{diag} (e^{j\theta_1} \dots, e^{j\theta_N}), \in \mathbb{C}^{(N\times N)}$ is the phase shift matrix at the IRS with $\theta_i \in [0, 2\pi]$, \textbf{G} $\in \mathbb{C}^{(M \times K)}$ denotes the beamforming matrix at the BS, $\textbf{x} \in \mathbb{c}^{(K\times 1)}$ signifies the transmitted signal with zero mean and $\mathbb{E}[ |x|^2] = 1$, and finally, $\omega_k$ indicates the zero mean additive white Gaussian noise (AWGN) with entries of variance $\sigma^2$.

Under actual restrictions, this letter considers maximum SE via simultaneous optimization of the beamforming matrix \textbf{G} and the phase shift matrix \textbf{$\Phi$}. The SE is given by \cite{ref8}, \cite{ref9} as

\begin{equation}
    R = \sum_{k = 1}^K \log_{2}(1 + \frac{|(\textbf{h}^T_{r,k}\Phi \textbf{H}_1 + \textbf{h}^T_{d,k})\textbf{g}_k|^2}{\begin{matrix} \sum_{i, i \ne k}^{K} |(\textbf{h}^T_{r,k}\Phi \textbf{H}_1 + \textbf{h}^T_{d,k})\textbf{g}_i|^2 + \sigma_{i}^2 \end{matrix}}),
\end{equation}
in which $\textbf{g}_k$ refers to the \emph{k}-th column of the \textbf{G} matrix.
The optimization problem can be formulated as follows

\begin{subequations} \label{Eq:OptPoutOMAdef}
\begin{align}
{\mathop {\max }\limits_{ \Phi,\textbf{G} } }&  \;\;R
\\
\text{s.t.} \;\; & \text{trace}{\{\textbf{GG}^H\}} \le P_t                          \label{Eq:OptPoutOMAdefC1}
\\
           & \Phi = \text{diag} (e^{j\theta_1}, \dots, e^{j\theta_N})
           \label{Eq:OptPoutOMAdefC2}
\end{align}
\end{subequations}

and $P_t$ denotes the total permitted transmission power.
\section{Hybrid Beamforming Design}
 This section provides an overview of the TD3 methodology.  While the principles of DRL and DDPG as well as their application in wireless communication in \cite{ref6} and \cite{ref10}, in the following, the mathematical principles in such a methodology will be further discussed for the sake of clarification. Moreover, after explaining the elements of the intended DRL, the deep neural network (DNN) will be described in depth.
\subsection{Overview of TD3}\label{AA}
TD3 algorithm \cite{ref11} is a model-free, online, off-policy reinforcement learning technique. The TD3 is an actor-critic reinforcement learning method that seeks out the greatest feasible line of action to maximize the anticipated long-term cumulative reward. DDPG agents might overestimate value functions, which can produce suboptimal policies.  Utilizing two critic networks is the first new feature for TD3. The method used in DRL with Double Q-learning, which involved calculating the current Q value using a second target value function to reduce the bias, served as an inspiration for this method. Furthermore, it postpones updating the actor network in order to overcome the overestimation. The critic networks continue to update after each time step while the actor network and target networks update after a certain number of time steps \cite{ref12}.

\begin{algorithm}
\caption{TD3 framework for hybrid beamforming optimization}

\begin{algorithmic}
\State \textbf{Input:} $\textbf{H}_1, \textbf{h}_{r,k}, \textbf{h}_{d,k}, \forall k$
\State\textbf{Output:}
    The current SE as the result of optimal action: a 
    $ = {\{\textbf{G}, \textbf{\text{$\Phi$}}\}}$
\State\textbf{Initialization:} 
    Both critic networks $Q_{\theta_1}$ , $Q_{\theta_2}$and actor network $\pi_\phi$ with random parameters $\theta_1$, $\theta_2$, $\phi$.
    Target networks with following procedure: $\theta_{1}^{'} \leftarrow  \theta_{1}, \theta_{2}^{'} \leftarrow  \theta_{2}, \phi^{'} \leftarrow  \phi$.
    Replay buffer $\mathcal{B}$,
    Beamforming matrix \textbf{G}, phases shift matrix $\Phi$
\begin{algorithmic}[1]
\For{\texttt{n = 0 to N-1}}
        \State Obtain the initial state $s^{(0)}$ using the current CSI $\text{  }\;\;\;$ $(\textbf{H}_1, \textbf{h}_{r,k}, \textbf{h}_{d,k})$
        \For{\texttt{t = 0 to T-1}}
    \State Select action $a^{(t)} = \{\textbf{G}^{(t)}, \Phi^{(t)}\} = \pi_{\phi}(s^{(t)})$
        \State Observe reward $r^{(t)}$ and new state $s^{(t+1)}$
        \State Store transition tuple $(s^{(t)}, a^{(t)}, r^{(t)}, s^{(t+1)})$
        \State Sample a $\mathcal{W}$ mini-batch $(s^{(t)}, a^{(t)}, r^{(t)}, s^{(t+1)})$  $\text{  }\;\;\;\;\;\;\;\;$ of replay buffer $\mathcal{B}$
        \State $\check{a} \leftarrow \pi_{\phi^{'}}(s^{(t+1)})$
    
        \State $y  \leftarrow r + \lambda min_{i=1,2} Q_{\theta_i^{'}}(s^{'}, \check{a}) $
        \State Update critics $\theta_i \leftarrow argmin_{\theta_i} \mathcal{W}^{-1}\sum (y - \text{  }\;\;\;\;\;\;\;\;\;\;Q_{\theta_i}(s,a))^{2}$
        \State \textbf{every U step:}
        \State Updating $\phi$ by the deterministic policy gradient: $\text{    }\;\;\;\;\;\;\nabla_\phi J(\phi) = \mathcal{W}^{-1}\sum\nabla_a Q_{\theta_1}(s,a)|_{a=\pi_{\phi}(s)}\nabla_\phi \pi_\phi(s) $
        \State Soft update target networks via (11) and (12)
        \EndFor
        \State $s^{(t)} \leftarrow s^{(t+1)}$
        \EndFor
\end{algorithmic}
\end{algorithmic}
\end{algorithm}

\subsection{Mathematical Details}\label{BB}
Reinforcement learning takes into consideration an agent's interaction with its environment to learn a behavior that maximizes rewards. At each discrete time step $t$, the agent chooses actions $a^{(t)} \in A$ based on its policy $\pi$: $S \rightarrow A$ earning a reward $r^{(t)}$ and the new state of the environment $s^{(t)}$. Return is defined as the discounted sum of rewards $R_t = \sum_{i=t}^T {\gamma^{i-t}r(s^{(i)}, a^{(i)})}$ where $\gamma$ is a discount factor determining the priority of short-term. After each time step $t$, the tuple $(s^{(t)}, a^{(t)}, r^{(t)}, s^{(t+1)})$ is stored in the replay buffer  $\mathcal{B}$ with size $\mathcal{D}$ for use in calculating the loss functions [10].
 
Reinforcement learning discovers the strategy $\pi_\phi$ that maximizes expected return, given parameters $\phi$. $Q_\theta (s,a)$ calculates the anticipated reward for action $a$ in state $s$ using the parameter $\theta$.
The method is described by Algorithm 1. TD3 simultaneously learns $Q_{\theta_1}$ and $Q_{\theta_2}$. $\pi_{\phi}(s^{(t)})$ with parameter $\phi$ at time step $t$ determines the selected action.
Additionally, three target networks are duplicated based on their originals. the neural networks are necessary for estimating the goal value and optimizing the actor network's output in the absence of its actual value.

The rest of the section will focus on the loss function's innermost region to show how TD3 works and how it differs from DDPG \cite{ref13}. By sampling $\mathcal{W}$ from  $\mathcal{B}$, the target action is obtained as follows
\begin{equation}
  a' = \pi_{\phi^{'}}(s^{(t+1)}).
\end{equation}
One target value is used for both Q-functions, calculated using whichever of the two Q-functions gives a smaller target value as
\begin{equation}
  y = r + \gamma  
\underset{i = 1,2}{\text{min}}
\  Q_{(\theta_{\text i})}(s^{(t+1)}, a^{'}).
\end{equation}
By sampling a mini-batch with the size of $\mathcal{W}$ and then both are learned by regressing to the following loss functions
\begin{equation}
L(\theta_1) = {E}[(Q_{\theta_1}(s^{(t)}, a^{(t)}) - y)^2],
\end{equation}
\begin{equation}
L(\theta_2) = {E}[(Q_{\theta_2}(s^{(t)}, a^{(t)}) - y)^2],
\end{equation}
and the parameters of the critic networks are updated with the following procedure
\begin{equation}
\theta_{i}^{(t+1)} = \theta_{i}^{t} - \mu_i\nabla_{\theta_i}L(\theta_i),
\end{equation}
in which $\mu_i$ shows the utilized learning rate for updating both Q functions.
After each $U$ iteration, the actor network and target networks will be updated as
\begin{equation}
J(\phi) = \nabla_a Q_{\theta_1}(s,a)|_{a=\pi_{\phi}(s)}\nabla_\phi \pi_\phi(s),
\end{equation}
\begin{equation}
\phi^{(t+1)} = \phi^{(t)} - \mu_a\nabla_\phi J(\phi),
\end{equation}
where $\mu_a$ is the updating learning rate for actor network. As it is obvious, only the gradient of the first Q-network is considered in the updating process.
The target critic network and the target actor network are  updated as follows
\begin{align}
\theta^{'} \leftarrow \tau_c\theta + (1 - \tau_c)\theta^{'}, \\ \phi_i^{'} \leftarrow  \tau_a\phi_i + (1 - \tau_a)\phi_i^{'},
\end{align}
respectively, where $\tau_c$ and $\tau_a$ are the learning rates for target networks respectively.
\subsection{Elements of the DRL Framwork}
First, the state, action and reward should be characterized for the proposed joint design of transmit beamforming and phase shifts.
To do so, they are characterized as follow \cite{ref6}
\begin{enumerate}
  \item The state set $\mathbb{S}(t)$ at the time step $t$ is determined by:
  \begin{itemize}
  \item The transmission power at the $t^{th}$ time step
  \item The received power of users at the $t^{th}$ time step,
  \item The action from the $(t-1)^{th}$ time step
  \item The channel matrix $\textbf{H}_1$ and $\textbf{h}_{k,r}, \textbf{h}_{k,d}, k \in [1, K]$
  \end{itemize}
  
  \item The action space is simply constructed by the transmit beamforming matrix $\textbf{G}$ and the phase shift matrix $\Phi$.
  
  \item	Reward in time step $t$ is defined as achieved SE based on given matrix \textbf{G}, matrix $\phi$, and the instantaneous channels $\textbf{H}_1$, $\textbf{h}_{k,r}$, $\textbf{h}_{k,d} \forall k$ 
\end{enumerate}
Since the input of a neural network cannot be a complex number, two neurons should be assigned to a complex number, one for the real component and one for the imaginary part.
\subsection{DNN Structure}\label{DD}
Fig.2 depicts the construction of the DNN used in the study. Both the actual and target networks are fully connected DNNs with input, hidden, and output layers.
Similar to the DDPG [6], the input of the critic network layer has the same dimension as the state and action sets. The actor network only uses the state set. Hidden network size is related to the number of users, the number of the BS antennas and the number of the IRS reflectors.
\emph{Tanh} is used as the activation function of the neurons because it covers negative inputs better. Adaptive ADAM optimization updates weights and biases with learning rate $\mu^{(t)} = \lambda \mu^{(t+1)}$, where $\lambda$ represents the training network decay rate.
\begin{figure}[htbp]
\centerline{\includegraphics[width=2.5 in]{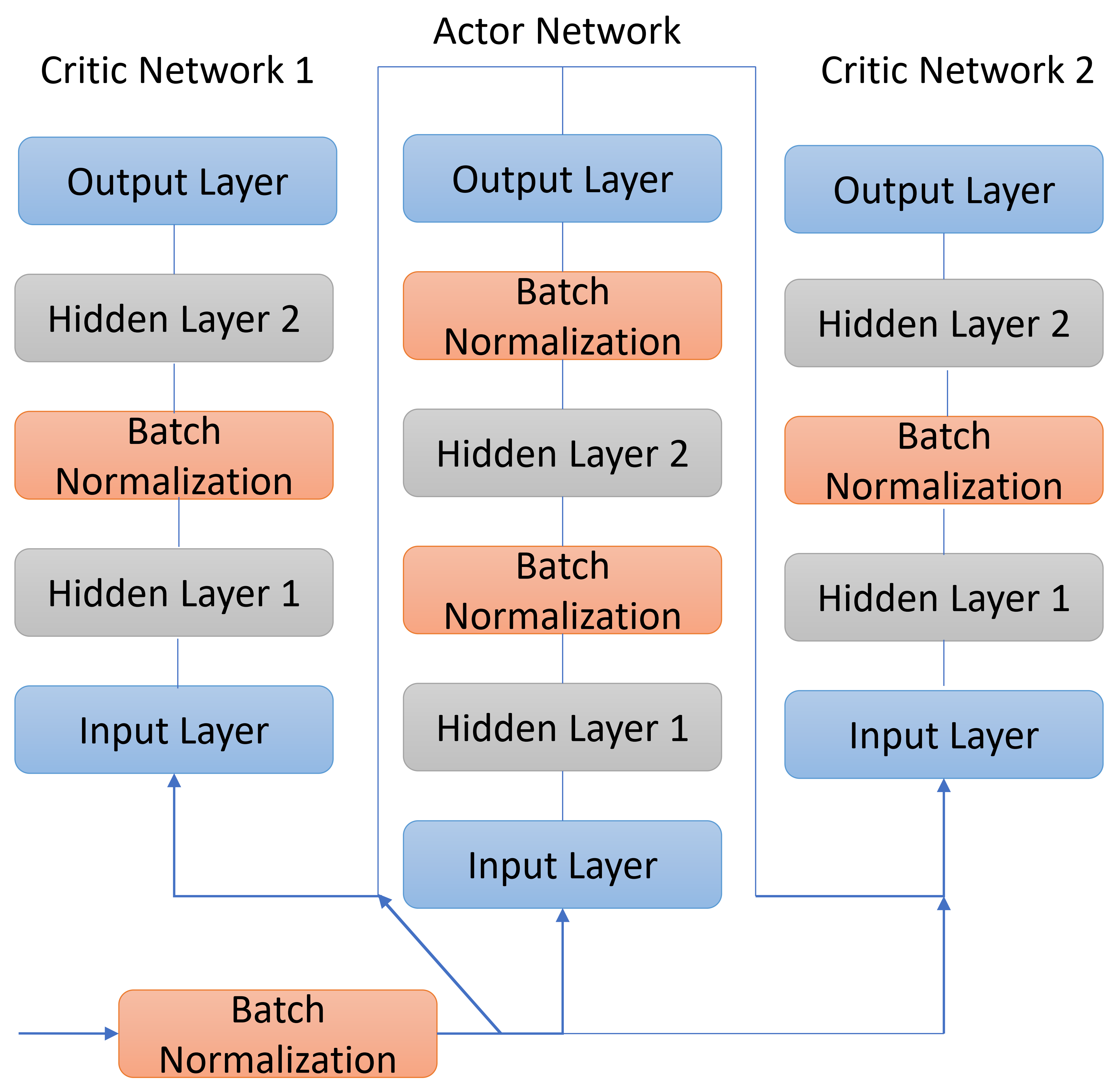}}
\caption{The structure of the utilized DNN.}
\label{fig}
\end{figure}
 
\section{Numerical Results}
In this section, the proposed method is evaluated in terms of the spectral efficiency. Also, the numerical results are compared against the Alternating Optimization method as a comparison benchmark. To do so, $\textbf{h}_{d,k} \in \mathbb{C}^{(M \times 1)}$ is modeled using a random Rayleigh distribution. Channels BS-IRS and IRS-users follow Rician fading. The BS-IRS channel is identified as \cite{ref14}
\begin{equation}
\textbf{H}_1 = \sqrt{\frac{K_1}{K_1 + 1}}\Bar{\textbf{H}}_1 + \sqrt{\frac{1}{K_1 + 1}}\tilde{\textbf{H}}_1,
\end{equation}
where $K_1$ represents the Rician K-factor of $\textbf{H}_1$, $\Bar{\textbf{H}}_1 \in \mathbb{C}^{(N \times M)}$ denotes the LoS component, which does not change during the channel's coherent time, and $\tilde{\textbf{H}}_{1} \in \mathbb{CN}^{(N \times M)}$.In parallel, the channel between the IRS and the \emph{k}th user is defined as follows

\begin{equation}
\textbf{h}_{r,k} = \sqrt{\frac{K_2}{K_2 + 1}}\Bar{\textbf{h}}_{r,k} + \sqrt{\frac{1}{K_2 + 1}}\tilde{\textbf{h}}_{r,k},
\end{equation}
in which $K_2$ denotes the Rician K-factor  of $\textbf{h}_{r,k}$, and $\Bar{\textbf{h}}_{r,k} \in \mathbb{C}^{(1 \times N)}$ stands for the LoS component, which remains stable during channel coherent time, and $\tilde{\textbf{h}}_{r,k} \in \mathbb{CN}^{(1 \times N)}{(0,1)} $ indicates the non-Los (NLoS) component.
$\Bar{\textbf{H}_1}$ and $\Bar{\textbf{h}}_{r,k}$ are described respectively as

\begin{equation}
\Bar{\textbf{H}}_1 = a_N^H(\theta_{AoA,1})a_M(\theta_{AoD,1})
\end{equation}
\begin{equation}
\Bar{\textbf{h}}_{r,k} = a_N(\theta_{AoD,2})
\end{equation}in which $a_N(\theta) = [1, \exp^{j2\pi\frac{d}{\lambda}\sin\theta}, \dots,  \exp^{j2\pi\frac{d}{\lambda}(N-1)\sin\theta}] $, and $AoD,1$ and $AoA,1$ indicate the signal's BS departure angle and IRS arrival angle. $AoD,2$ presents the IRS-user's angle of departure.
We assume the learning and decay rates are $10^{-3}$ and $10^{-5}$, and the discount factor is $0.99$. Moreover, the BS and IRS are fixed at a horizontal distance of 51 meters, while their vertical height with the users is 2 meters [8]. The users are randomly placed between the BS and IRS independently in each iteration, and $U$ and the maximum number of iterations are 1 and 8000 respectively. The SE is the largest network reward found, and $K = N = M = 4, K_1 = K_2 = 10$, and $P_t = 30$ dB.
Fig. 3 compares the the performance of the methods by evaluating the impacts of the allocated power on the SE. As expected, the TD3 algorithm leads to a better result than DDPG which \cite{ref6} proves its efficiency, and the alternative optimization approach in \cite{ref15}, a baseline method and benchmark for many articles. It is apparent that TD3 is more practical, especially in low powers.
\begin{figure}[htbp]
\centerline{\includegraphics[width=0.7\columnwidth]{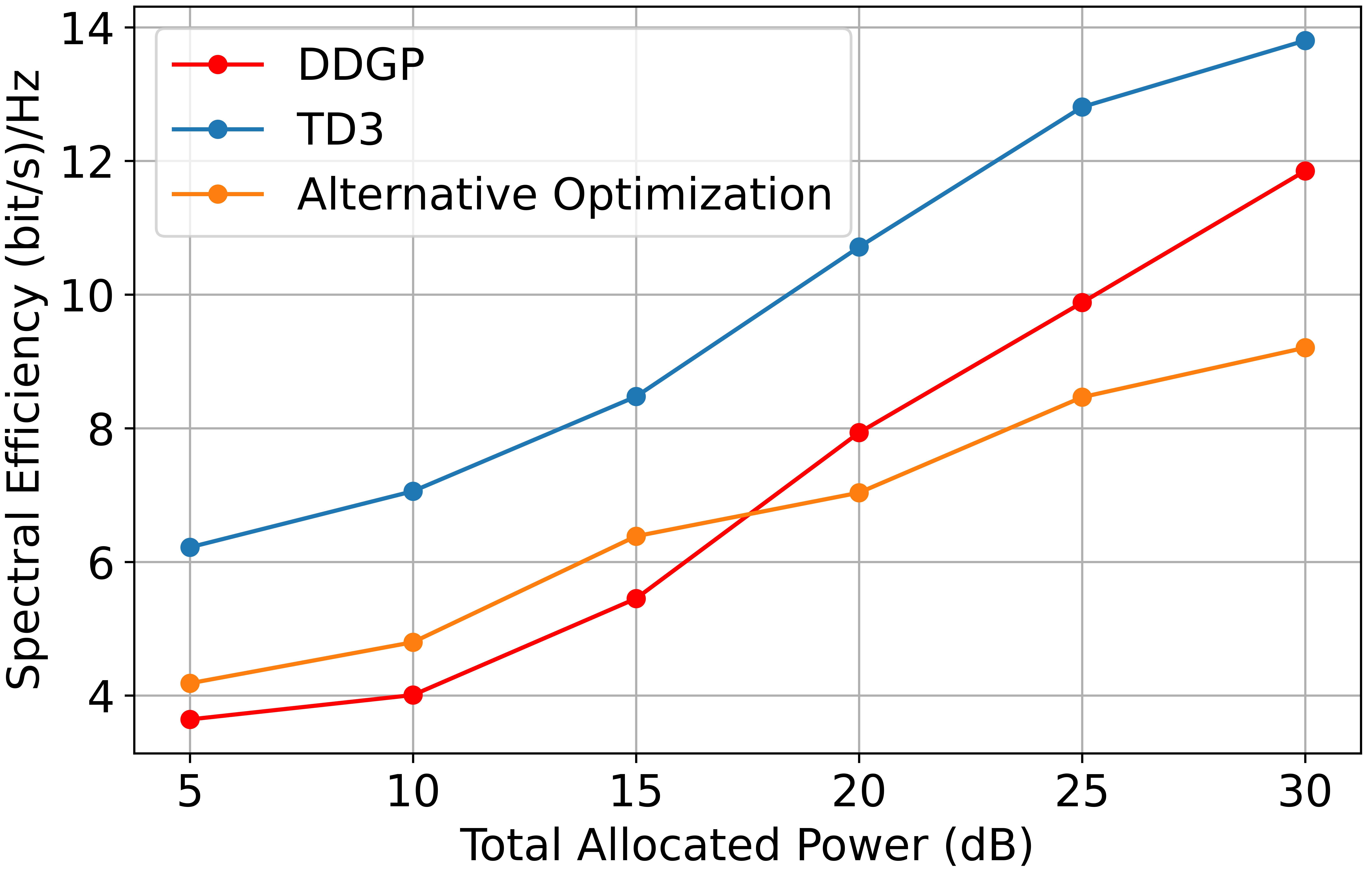}}
\caption{Spectral efficiency obtained with various allocated power.}
\label{fig}
\end{figure}
Finally, Fig. 4 illustrates the average rewards obtained based on $N$ and iterations. As it is evident, increasing the number of $N$ and iterations will cause the average reward to rise.
\begin{figure}[htbp]
\centerline{\includegraphics[width=0.7\columnwidth]{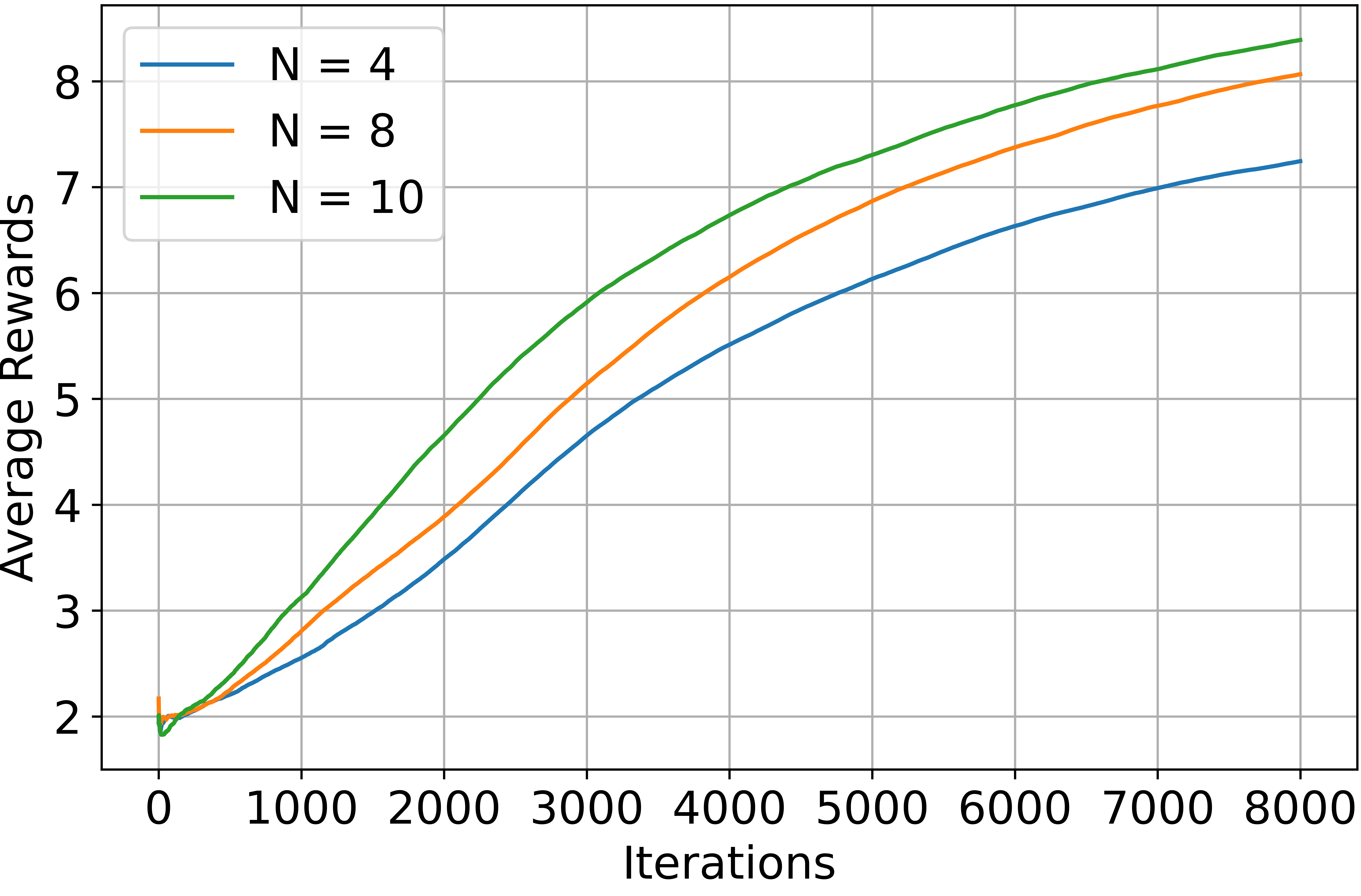}}
\caption{The average rewards for different numbers of reflecting element (N).}
\label{fig}
\end{figure}
\section{Complexity Analisys}
The criteria employed for comparing the DDPG and TD3 are included the number of trainable parameters, the required size for saving the entire neural networks, and the spent time on the training process. Table 1 compares the mentioned factors in details based on the mentioned hyperparameters. As it is evident, TD3 is more complicated than DDPG. However, it presents a better performance in return.
\begin{table}
\begin{center}
\caption{Comparing the complexity of DDPG and TD3.}
\label{tab1}
\begin{tabular}{ |c|c|c| } 
 \hline
     & DDPG & TD3\\ 
\hline
  Number of Trainable Parameters & $4.56\times10^{5}$ & $7.29\times10^{5}$ \\
\hline
Memory Usage & 968 KB & 1.41MB \\ 
\hline
Each Episode Duration & 79.88s & 99.85s \\ 
\hline
\end{tabular}
\end{center}
\end{table}
\section{Conclusion}
In this letter, we introduced the TD3 DRL method, which jointly optimizes the active beamforming matrix at the BS and the passive phase shift matrix at the IRS to enhance the SE in a multiuser MISO system. While the majority of the earlier research used alternative optimization techniques to achieve this goal, it was one of the first works that investigated the application of DRL in terms of the join optimization problem. The main purpose of this study was showing the superiority of the TD3 over the DDPG which has been revealed through the numerical results.

\end{document}